# Realtime Particulate Matter and Bacteria Analysis of Peritoneal Dialysis Fluid using Digital Inline Holography

Nicholas Bravo-Frank[1, 2], Nicolas Mesyngier[2], Lei Feng[2], and Jiarong Hong [1, 2, 3, 4]

1: Department of Electrical and Computer Engineering, University of Minnesota
2: Saint Anthony Falls Laboratory, University of Minnesota
3: Minnesota Robotics Institute, University of Minnesota
4: Department of Mechanical Engineering, University of Minnesota
**Corresponding Author E-mail: jhong@umn.edu**

**Abstract**

We developed a digital inline holography (DIH) system integrated with deep learning algorithms for real-time detection of particulate matter (PM) and bacterial contamination in peritoneal dialysis (PD) fluids. The system comprises a microfluidic sample delivery module and a DIH imaging module that captures holograms using a pulsed laser and a digital camera with a 40× objective. Our data processing pipeline enhances holograms, reconstructs images, and employs a YOLOv8n-based deep learning model for particle identification and classification, trained on labeled holograms of generic PD particles, *Escherichia coli* (*E. coli*), and *Pseudomonas aeruginosa* (*P. aeruginosa*). The system effectively detected and classified generic particles in sterile PD fluids, revealing diverse morphologies predominantly sized 1–5 μm with an average concentration of 61 particles/μL. In PD fluid samples spiked with high concentrations of *E. coli* and *P. aeruginosa*, our system achieved high sensitivity (>90%) in detecting and classifying these bacteria at clinically relevant low false positive rates (~0.5%). Further validation against standard colony-forming unit (CFU) methods using PD fluid spiked with bacterial concentrations from approximately 100 to 10,000 bacteria/mL demonstrated a clear one-to-one correspondence between our measurements and CFU counts. Our DIH system provides a rapid, accurate alternative to traditional culture-based methods for assessing bacterial contamination in PD fluids. By enabling real-time sterility monitoring, it can significantly improve patient outcomes in PD treatment, facilitate point-of-care fluid production, reduce logistical challenges, and be extended to quality control in pharmaceutical production.

# 1. Introduction

End-stage kidney disease (ESKD) is the final stage of chronic kidney disease (CKD), where the kidneys have lost nearly all their ability to function effectively, necessitating life-sustaining treatment. In the United States alone, more than one in seven adults suffers from CKD, with over 808,000 individuals living with ESKD (USRDS, 2022). The management of ESKD predominantly involves dialysis, with approximately 69% of ESKD patients undergoing this treatment, while the remainder receive a kidney transplant (USRDS, 2022). Dialysis is available in two forms: hemodialysis (HD) and peritoneal dialysis (PD) (Kaplan, 2017; Foo et al., 2020). While HD requires a dialyzer to filter blood and is typically performed in a hospital or clinic setting, PD utilizes the peritoneum as a natural filter, offering the convenience of home treatment (Hechanova, 2022a; Hechanova, 2022b). PD is particularly advantageous due to its accessibility, enabling patients to manage their treatment independently and at home, with exchanges lasting around 30 minutes and the solution retained for 4-6 hours (Li et al., 2017; Milan et al., 2020; Chen et al., 2021; Paudel et al., 2021).

However, PD entails significant environmental and logistical challenges, primarily due to the large volumes of PD fluid required—approximately 2-3 liters per treatment, used 4-6 times daily, amounting to around 300 liters per month per patient. This fluid is commonly produced in centralized facilities and then distributed to patients, a process that incurs substantial transportation costs and environmental impacts (Yau et al., 2021; Yeo et al., 2022), including increased greenhouse gas emissions due to the energy-intensive nature of manufacturing and logistics. A study by Chen et al. (2017) highlighted that the annual carbon footprint per PD patient in China reaches approximately 3.4 tons of $CO_2$ equivalents, with the production and transportation of PD fluids accounting for a significant portion of these emissions. This centralized production and distribution model also raises the cost of PD, as importing sterile PD fluid often makes it more expensive than HD in developing countries, while also increasing vulnerability to crises such as the COVID-19 pandemic (Paudel et al., 2021; Jeloka et al., 2020; Chen et al., 2021). The concept of generating PD fluid at the point of care, directly in patients' homes, emerges as a potential solution to these issues, especially in low-resource regions. This strategy could substantially lower costs and environmental impact, while improving patient quality of life by reducing the need for large storage volumes of PD fluid and dependence on complex supply chains for this time-sensitive, life-saving treatment.

The feasibility of in-home PD fluid production hinges on stringent quality control measures, particularly concerning particulate matter (PM) and sterility standards. Current methods for PM analysis in PD fluids are largely driven by United States Pharmacopeia (USP) <788> (USP, 2012), which specifies standards for particulate matter in liquids. The primary approach for PM analysis involves light obscuration and microflow imaging (MFI). While these methods are standard, they are not optimized for point-of-use production and can exhibit significant deviations from actual concentrations, with expected deviations ranging from 25-50% for protein aggregates (Narhi et al., 2015; Corvari et al., 2015; Kiyoshi et al., 2019; Fawaz et al., 2023). Furthermore, MFI, despite offering detailed morphological information and being capable of detecting a broad size and concentration range of particles, is prohibitively expensive for routine use in point-of-care manufacturing quality assurance. It also exhibits variability in sample loading, which affects the consistency of the analysis (Fawaz et al., 2023).

Sterility testing, as outlined in USP <71> (USP, 2008), mandates traditional culturing methods. While these methods are considered the gold standard, they are too time-consuming for immediate, point-of-use generation, often requiring up to 14 days to produce results with the expertise of a trained microbiologist or technician (Pitt et al., 2012; Tien et al., 2020). Emerging rapid microbiological techniques such as Polymerase Chain Reaction (PCR) and Mass Spectrometry (MS) have revolutionized microbial detection due to their high specificity and sensitivity (Kim et al., 2012; Lin et al., 2016; Kabiraz et al., 2023). However, these approaches are hindered by the need for specialized equipment and expertise, rendering them impractical for home-based PD fluid production (Panwar et al., 2023). Additionally, immunological methods like Enzyme-Linked Immunosorbent Assay (ELISA) and lateral flow immunoassays offer rapid on-site testing but suffer from variable sensitivity and high rates of false positives (Panwar et al., 2023). Innovative technologies like biosensors (Huo et al., 2021; Wang et al., 2022) offer real-time, portable solutions but still face issues related to sensitivity, specificity, and reliability, preventing them from fully meeting industrial standards. Some label-free technologies, including Fourier Transform Infrared spectroscopy (Bâcioğlu et al., 2019), hyperspectral imaging (Kang et al., 2020), and quantitative phase imaging (Choi et al., 2021), are primarily used in research laboratories due to their computational demands, operational complexity, and high costs. Additionally, both culture-based and many rapid methods struggle to detect low-concentration pathogens, which are commonly found in sterile liquids in pharmaceutical products, including PD fluid.

In this context, Digital Inline Holography (DIH) offers a cost-effective and compact solution for the high-throughput analysis of microparticles in suspension (Katz et al., 2010; Berg, 2022). By utilizing a digital camera, DIH captures interference patterns—holograms—created when coherent light, typically from a laser, scatters as it passes through a particle suspension. The unscattered portion of the light then interferes with the scattered portion, creating a hologram. Holograms encode essential information about the particles' three-dimensional positions, morphologies, and refractive indices which are associated with the biophysical properties. These can be extracted through numerical reconstruction. Combined with deep learning approach, DIH offers far more detailed insights than conventional brightfield microscopy, including biochemical composition, cell viability, and metabolic states without the need for reagents or fluorescent labels (Guo et al. 2021; Gul et al., 2021; Zeng et al., 2021; Martin et al., 2023; Sanborn et al., 2023; Barua et al. 2023), enabling in-depth analysis of both generic PM and bacteria. DIH also offers a depth of field approximately 1000 times greater than that of conventional microscopy (Katz et al., 2010) which significantly increases throughput. However, despite its demonstrated versatility in various PM-related applications (Alexander et al., 2020; Luo et al., 2021; Martin et al., 2022), its broad application for tasks like monitoring the sterility of PD fluids in home-based production is hindered by the traditional numerical approaches used for hologram processing and inference, being both computationally, and resource intensive. For DIH to be effective in a point-of-care setting, the system must offer real-time, autonomous operation, delivering fast and accurate results without requiring highly trained personnel.

To bridge this critical technological gap, we have developed a user-friendly and cost-effective DIH sensor system tailored for the precise analysis of particulate matter and bacteria in peritoneal dialysis fluids. By integrating advanced deep learning algorithms, our system provides real-time, autonomous analysis without the need for specialized personnel. This innovation has the potential to facilitate portable, in-home PD care, significantly reducing the risk of bacterial infections and potentially transforming the management of ESKD. This paper is organized as follows: Section 2 describes our custom-built DIH imaging system and the materials used, detailing technical specifications and experimental setup. Section 3 presents the application of our system in analyzing particulate matter in sterile PD fluids and validates its performance using PD fluid samples spiked with two types of bacteria, comparing the results to colony-forming unit (CFU) counts. Finally, Section 4 discusses our findings and their implications, highlighting the potential impact on patient care and the broader field.

## 2. Materials and Methods

To address the need for real-time detection of PM and bacteria in PD solutions, we developed a DIH imaging system integrated with a customized deep learning model. The system analyzes holographic images of samples flowing through a microfluidic channel, detecting, and classifying particles, as well as determining the concentration and size distribution.

### 2.1. Hardware

The hardware configuration of the system, illustrated in Figure 1, comprises a DIH imaging module and a sample delivery module. The DIH imaging module employs a low-cost finite conjugate microscope setup. It utilizes a 405 nm pulsed laser with a 1 µs pulse width to illuminate the sample. Image capture is performed by a 1.6-megapixel CMOS camera (FLIR BFS-U3-16S2M-CS), operating at 100 frames per second (fps) through a 40× objective lens. The resultant setup provides a field of view (FOV) of 124 µm (length) × 93 µm (width) with a spatial resolution of 87 nm/pixel. The key component of the sample delivery module is a custom-made two-inlet polydimethylsiloxane (PDMS) microfluidic chip. One inlet connects to the sample channel, and the other connects to two sheath flow channels. The sample channel has a cross-section measuring 200 µm (width) × 80 µm (depth), and the sheath flow enters at a 45-degree angle to precisely focus the sample stream. Both the sample and sheath flows are driven by a customized pressure pump, operating at rates of approximately 3 µL/min and 3.6 µL/min (controlled with pressure regulators), respectively. This narrows the sample flow stream to a width of 90 µm, fitting within the FOV of our sensor. During imaging, the focal plane is set to the center of the channel depth, using the channel walls as a reference.

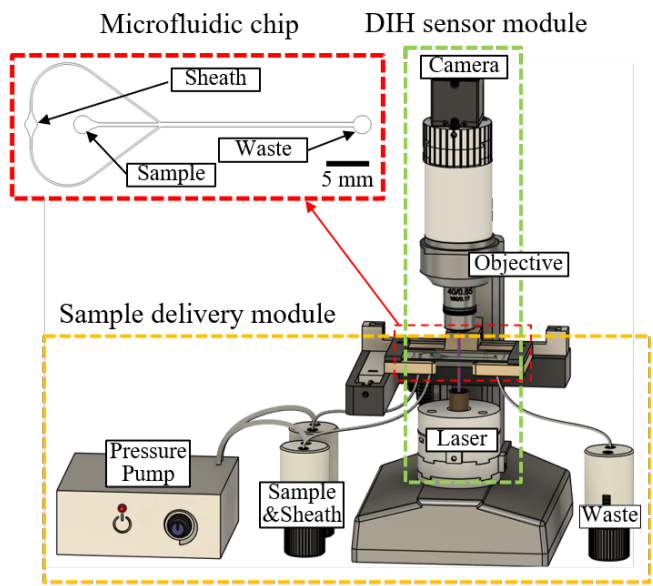

**Figure 1.** Illustration of digital inline holography (DIH) imaging system for peritoneal dialysis fluid analysis including the DIH sensor module and sample delivery module. The inset figure shows the microfluidic chip used in the sample delivery module.

## 2.2. Software

We developed a graphical user interface (GUI) to operate both the DIH imaging and sampling delivery modules, providing a comprehensive interface for managing various system functions. The software features control panels for managing sample delivery (e.g., flow rate), laser parameters (e.g., energy, pulse frequency, and pulse width), and camera settings (e.g., exposure time, frame rate, etc.) for hologram capture to ensure optimal image quality. It also enables real-time visualization of data and analysis results, allowing for immediate feedback and adjustments during experiments.

Experimental data were captured at 100 fps, pulse width of 1 µs, sample flow rate of 3 µL/min, and 40× magnification. Two bacterial species, *Escherichia coli* (*E. coli* MG1655) and *Pseudomonas aeruginosa* (*P. aeruginosa*, PAO1 JCM 14847), were prepared using the ATCC Bacteriology Culture Guide (ATCC, 2022). Cultures were revived from frozen stocks and grown in Luria-Bertani (LB) broth medium at 37 °C for 24 hours. The initial concentration was estimated by measuring optical density at 600 nm ($OD_{600}$). The original culture was serially diluted into sterile PD fluid to achieve final concentrations of $10^6$ bacteria/mL for imaging. The PD fluid was filtered through a 0.2 µm membrane to remove excess PDP particles to ensure accurate labeling of spiked particles. The two separate solutions, *E. coli*–spiked PD fluid and plain PD fluid, were run through our system to create a dataset. All preparation and imaging procedures were conducted within a Biosafety Level 2 (BSL-2) biosafety cabinet to avoid contamination. For each sample, a total of 1 mL of fluid was imaged. Samples were collected using sterile Corning microtubes and stored on ice to prevent additional bacterial growth prior to plating. The dataset was then manually labeled to create an initial training set, consisting of approximately 1,000 PD particles and 500 *E. coli* images, which were used to train a detection model for all particles, generating bounding boxes around detected particles. All particles were subsequently classified manually based on the known sample and using reconstruction to identify ambiguous particles. Bacterial plating was performed on LB agar plates (BD 244520) to quantify CFUs. The volume of liquid plated varied based on the expected bacterial concentration, aiming for the recommended 25–250 colonies per plate. Plates were counted after incubation overnight at 30 °C.

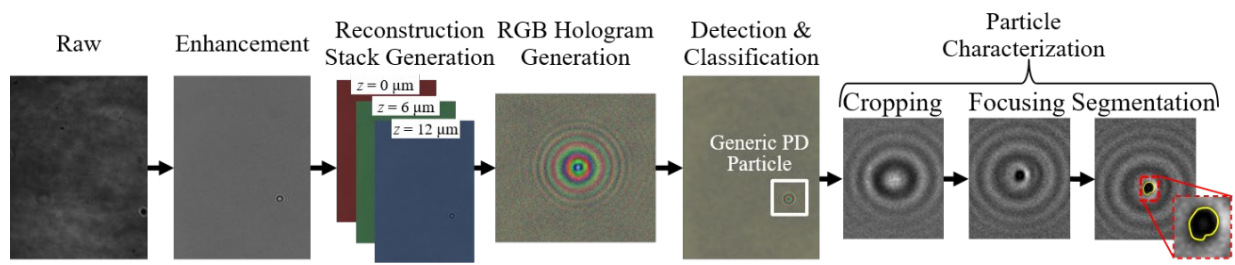

**Figure 2.** Schematics illustrating the hologram data processing framework which involves a series of steps including enhancement, reconstruction → RGB hologram, object detection and cropping, object focusing as well as object segmentation.

At the core of the software is the DIH processing pipeline, outlined in Figure 2. The pipeline encompasses six steps: enhancement, reconstructed stack generation, deep learning detection and classification, cropping of detections, focusing, and segmentation. Initially, raw holograms undergo enhancement through a fixed-size moving window subtraction of 40 frames, which removes static background noise and normalizes the image intensity for consistent image contrast. Subsequently, these enhanced images are reconstructed into additional focal planes at z = 6 µm and 12 µm relative to the imaging plane (centered in the channel depth), using a Fresnel diffraction kernel. These two reconstructed images are merged with the original image to create an RGB hologram that enhances hologram information for more precise classification outcomes.

This RGB hologram is then fed into a deep learning model utilizing the YOLOv8n architecture for particle identification and classification within bounding boxes. YOLOv8n is used for its memory efficiency and speed (Jocher et al., 2023). These bounding boxes facilitate the cropping of the enhanced grayscale images, with the model trained to ensure that individual particles are isolated with at least three diffraction rings maintained for reliable reconstruction. Following this, a Tenengrad focus metric, calculated as a summed Sobel gradient (Wu, 2015), is applied to determine the focus plane, assuming the in-focus particle has the sharpest edges. Lastly, segmentation is executed using the Meta Segment Anything Model (SAM) (Kirillov, 2023). We deploy SAM with a strategic point-based selection and a slight 3-pixel Gaussian blur to create a circle of five selection points around the image center, guaranteeing accurate segmentation even if the particle is off-center. An equivalent diameter calculated from the segmented area was used as the quantitative measure of particle size.

## 2.3. Deep Learning Model

The YOLOv8n model was trained on a dataset of 12,000 labeled holographic images of three types of particles in sterile PD fluid, with 4,000 images of each type. The particles include generic PD particles (PDP), *E. coli* (EC), representing a relatively small bacterium with peritrichous flagella, and *P. aeruginosa* (PA), representing a similar bacterium with polar flagella. The initial experimental data consisted of 2,000 images each for PDP, EC, and PA. These images were augmented to reach the final dataset through a combination of rotations, horizontal flips, vertical flips, contrast adjustments, cropping, and resizing. We allocated 70% of the data for training and 30% for validation. Training iterations were performed over 10 cycles as follows: hard data mining augmentation, where we systematically augmented the current subset of training images that had the lowest confidence scores or incorrect predictions, followed by 300 epochs of training using the base YOLOv8 training schedule. We then selected and saved the best model from the newest iteration and all past iterations and began another training iteration at the initial learning rate with this selected model.

## 3. Results

In this section, we validated our system and model using both sterile PD fluid and PD fluid spiked with EC or PA. Initially, we quantified the PDPs in the original sterile PD fluid, examining their concentrations, morphologies, and size distributions, and tested the detection capabilities of our deep learning model. We then evaluated the model's performance in classifying datasets labeled as PDPs versus EC and PDPs versus PA by spiking the PD fluid with either EC or PA. Finally, we assessed the sensitivity of our method through comparison with a culture-based CFU counting method.

### 3.1. Detection Assessment and Quantification of Particulate Matter in PD Fluid

In the first experiment, we used the DIH system to capture images of and assess the PDPs present in sterile PD fluid directly from the manufacturing package. Images were acquired at 50 fps with 40× magnification and a flow rate of 0.1 µL/min to ensure high-quality capture of PDPs with varying sizes and shapes. Figure 3(a) demonstrates the detection of PDPs in an enhanced hologram using our model, which locates each particle with a bounding box and provides a confidence score. A total of 5,000 individual PDPs, which were fully within the image frame without overlapping holographic signatures, were selected for further analysis.

Figure 3(b) shows a gallery of in-focus images of PDP particles reconstructed from cropped holograms in the original images. We observed a range of particle morphologies, including spheres, rods, and agglomerates, with sizes ranging from approximately 1–5 µm, as shown in Figure 3(c). Our system estimated an average concentration of PDPs of 61 particles/µL in the sterile prepackaged PD fluid. Figure 3(c) illustrates the distribution of the equivalent diameters of 10,000 detected particles. The histogram indicates that the particles have a mean size of 1.8 µm and a median size of 1.5 µm, with size ranges from 0.5 µm to 5 µm respectively. Particles below 1 µm, however, are difficult to size accurately due to the diffraction limit. Particles greater than 5 µm were detected, but extremely infrequently. This size range overlaps with that of some common bacteria, such as our tested EC and PA, making it necessary to differentiate these pathogens from the common PDPs.

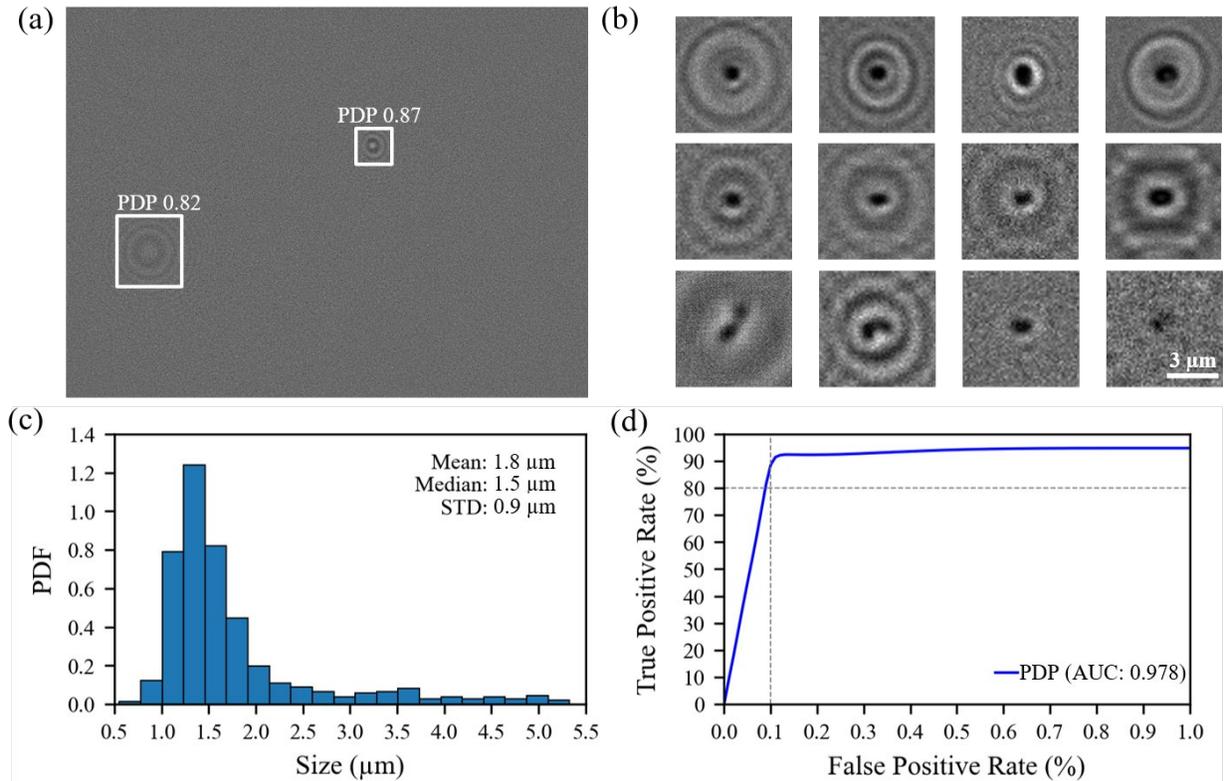

**Figure 3.** (a) An enhanced sample image showing detections of particles in peritoneal dialysis (PD) fluid using our deep learning model. Each detected particle is marked by a bounding box and labeled as PDP with the corresponding confidence score of its detection. (b) Samples of reconstructed PDPs showing a variety of size and morphology. (c) The histogram showing the probability density function (PDF) of the equivalent diameter of detected PDPs using our system. (d) Receiver operator characteristic (ROC) curve for detecting particles in PD fluid. Note that the vertical dashed line indicates our operating point of 0.1% false positive rate (FPR).

To assess our detection ability, we used the receiver operating characteristic (ROC) curve, shown in Figure 3(d), by comparing our true positive rate (TPR) against our false positive rate (FPR). The TPR, calculated as TPR = True Positives / (True Positives + False Negatives), represents the ability to correctly detect a particle, while the FPR, calculated as FPR = False Positives / (False Positives + True Negatives), represents the probability of incorrectly detecting background noise as a particle. Our ROC curve shows a high area under the curve (AUC) of 0.978, a metric which represents the likelihood of a correct prediction, demonstrating the robustness of our model in particle detection with a high rejection of noise. The ROC curve also illustrates the trade-off between TPR and FPR across different classification thresholds. At an operating point requiring 90% TPR our model has just a 0.1% FPR. Beyond 90% TPR, the curve plateaus with only gradual improvement, likely due to particles being occluded, either by the edge of the field of view or by other particles.

### 3.2. Model Assessment Using PD Fluid Spiked with Bacteria

To evaluate our model's classification accuracy, we conducted experiments by spiking PD fluid with high concentrations (>$10^6$ bacteria/mL) of EC or PA. For each condition: sterile PD fluid, EC spiked PD fluid, and PA spiked PD fluid, a total of 1 mL of sample was imaged, capturing holograms containing either only PDPs, PDPs with EC, or PDPs with PA. The ground truth was created by pre-labeled using our deep learning model, followed by manual checking of 2000 images for each bacteria spiked dataset. To balance the number of PDPs and bacterium, we supplemented 2000 PDP only images to each dataset. These two datasets, one PDP with EC, and one PDP with PA, were used to assess our model's performance. As shown in Figure 4(a), the RGB hologram of each particle type visually demonstrate qualitative differences in diffraction patterns among PDPs, EC, and PA, arising from variations in their optical properties, size, and morphology.

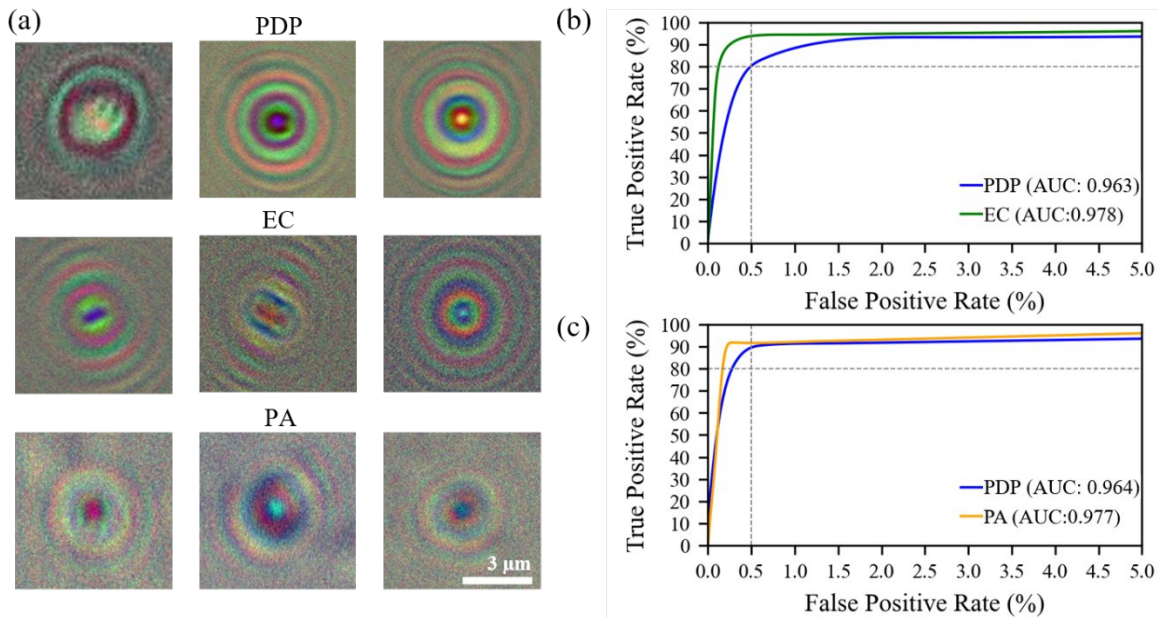

**Figure 4.** (a) Hologram samples showing RGB hologram for generic particles in peritoneal dialysis (PD) fluid (labelled as PDP in the figure), *E. coli* (EC) and *P. aeruginosa* (PA). Receiver operator characteristic (ROC) curves for (b) PDP and EC detections and for (c) PDP and PA with vertical dashed lines marking our operating point 0.5% false positive rate (FPR), given a minimum of 80% true positive rate (TPR).

Moreover, We quantitatively evaluated the model's ability to distinguish PDPs from bacteria using ROC curves. Figures 4(b) and 4(c) depict the model's classification performance for PDP vs. EC and PDP vs. PA, respectively. The high AUCs, approaching 0.98 for both EC and PA classifications, reflect the model's robustness in differentiating between PDPs and bacteria, even when particle sizes overlap. For instance, PDPs and EC share size ranges, with EC typically between 1–2 µm and PA slightly larger at 1.5–3 µm. Despite this overlap, the model effectively distinguishes the bacterial species from PDPs. We selected an FPR of 0.5% as it offers an optimal balance between high detection rates and minimal false positives—crucial for reliable classification in clinical settings. At this FPR, the TPR for detecting PA (Figure 4c) reaches approximately 90%, while for EC (Figure 4b), it is around 80%. This slightly lower TPR for EC is likely due to its smaller size compared to PA, leading to a higher rate of misclassification between EC and PDPs, considering the PDP mean size is around 1.8 µm. A detailed examination of misclassifications reveals that most errors occur when particles are occluded or when rod-shaped bacteria are oriented end-on, causing their diffraction patterns to resemble spherical PDPs. This effect is more pronounced in EC than in PA due to EC's smaller and more uniform shape, increasing the likelihood of misinterpretation in these orientations. Nevertheless, even in these challenging cases, the model maintains a high degree of accuracy, demonstrating its capability to handle variations in particle morphology and size. In summary, the ROC analysis demonstrates that our model effectively differentiates between PDPs and bacteria, even when particle sizes overlap. The system's performance, particularly at low FPRs, indicates its suitability for point-of-care settings, where minimizing false positives is crucial for ensuring patient safety and reliable monitoring.

### 3.3. Validation using culture-based method

To validate our model's ability to perform sterility testing compared to standard CFU methods, we conducted experiments using PD fluid spiked with varying concentrations of bacteria. Specifically, we prepared PD fluid samples containing either EC or PA at concentrations ranging from approximately 100 to 10,000 bacteria/mL. For each sample, we imaged approximately 400 µL of fluid using our DIH system to perform bacterial counting. We set a confidence threshold corresponding to a 0.1% FPR, as indicated in the ROC curves in Figure 4, which ensured a TPR of 80% for bacterial detection. While this threshold reduces the TPR for PDPs, in sterility detection we prioritize higher confidence in bacterial detection over perfect characterization of PDPs. The imaged fluid was collected and plated for CFU counts. To ensure accurate CFU enumeration, we used between 1 and 10 agar plates per sample, adjusting dilutions to maintain between 10 and 200 colonies per plate in accordance with standard microbiological practices. Our CFU counts confirmed that the samples used for DIH imaging had bacterial concentrations ranging from 36 CFUs/mL to 17,000 CFUs/mL, aligning with the concentrations used in our validation experiments.

Figure 5 illustrates the DIH bacterial counts plotted against CFU counts for each sample, with uncertainties for CFU estimated referencing Jongenburger at al. (2010), based on our colony counts, and for bacterial counts using a combination of our models FPR of 0.1% and the flow rate variations of our setup (10-20%), each represented as horizontal and vertical error bars respectively. The 45-degree line (red solid line) from the origin indicates a perfect 1:1 correlation between DIH counts and CFU counts. As shown, the data points closely align with this line across the tested bacterial concentration range, deviating by less than 10%. This deviation boundary is marked by a cone region formed by two additional straight lines (red dashed lines) around the 45-degree line, representing a ±10% deviation from perfect correlation. The alignment within this cone demonstrates that the deviations fall within the typical plating uncertainty reported by Jongenburger et al. (2010), which ranges from 5% to 20% depending on the total CFU count. These findings confirm a strong correspondence between bacterial concentrations determined by our DIH system and CFU counts over the tested range, where each CFU is assumed to originate from a single bacterium. Linear regression analysis yielded a high coefficient of determination ($R^2$) of 0.96, indicating a strong linear relationship between the two methods (black dashed line).

Despite the strong overall correlation, the regression line deviates from the 1:1 correlation below 61 counts/mL (indicated by the red dashed dotted horizontal line in Figure 5), reflecting reduced accuracy at lower concentrations due to the increased impact of false positives relative to true positives. At these low bacterial concentrations, the number of false positives—PDPs misclassified as bacteria—can become comparable to or exceed the number of true bacterial detections, compromising measurement reliability. As established in Section 3.1, the typical PDP concentration is about 61 particles/μL, and with our model operating at a 0.1% FPR, this translates to approximately 61 false positives/mL, defining our noise floor. At higher concentrations, extrapolation suggests potential deviations from the 1:1 correlation around 100,000 CFUs/mL or more. This may be due to the breakdown of the assumption that each bacterium yields one CFU, as well as challenges such as overlapping particles and model saturation in the DIH system. At elevated bacterial concentrations, multiple bacteria may occupy the same imaging area, resulting in overlapping holographic signatures that are difficult to resolve individually. Additionally, the processing capabilities of the model may become saturated, reducing the true positive rate and causing undercounting in DIH counts compared to CFU counts.

Nevertheless, within the experimental range of bacterial concentrations, i.e., above the noise threshold and below levels where CFU counting saturates or DIH undercounting occurs, our results demonstrate a strong correlation between the DIH method and standard CFU counts. This indicates that our DIH system is effective for sterility testing within this range, offering rapid and reliable results compared to traditional culture-based methods.

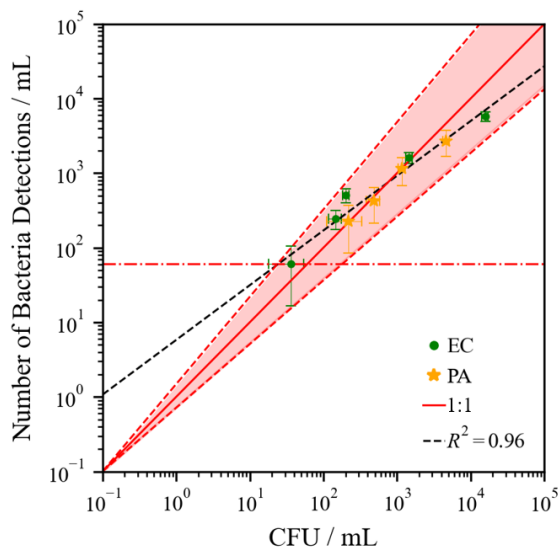

**Figure 5.** Correlation between bacterial counts from the DIH system and colony-forming unit (CFU) counts from traditional plating methods for peritoneal dialysis (PD) fluid samples spiked with *E. coli* (EC) or *P. aeruginosa* (PA) at concentrations ranging from approximately 36 CFU/mL to 17,000 CFU/mL. The results are presented for our deep learning model operating at 0.1% false positive rate (FPR). Data points plot DIH bacterial counts against CFU counts, with vertical error bars indicating uncertainties due to the model's FPR and flow rate variations (10–20%), and horizontal error bars representing CFU uncertainties based on Jongenburger et al. (2010). The solid red 45-degree line represents perfect 1:1 correlation with the two additional red dashed lines representing a ±10% deviation from perfect correlation. The red dashed-dotted horizontal line at 61 counts/mL denotes the system's noise floor, determined by the PDP concentration (61 particles/μL) and the 0.1% FPR. The black dashed line indicates the linear regression of the DIH bacterial counts against CFU counts.

## 4. Conclusion and Discussion

In this study, we developed a digital inline holography (DIH) system integrated with deep learning algorithms for real-time detection and analysis of particulate matter (PM) and bacterial contamination in peritoneal dialysis (PD) fluids. This system addresses the critical need for efficient sterility monitoring, enabling point-of-care fluid production and reducing logistical challenges associated with PD treatment. The system consists of a sample delivery module that ensures consistent and precise introduction of samples through controlled flow, and a DIH imaging module that captures high-quality holographic images using a pulsed laser and a CMOS camera equipped with 40× microscope objectives. Our data processing pipeline enhances raw holograms, reconstructs images at multiple focal planes, and employs a deep learning model for particle identification and classification. Specifically, the deep learning model utilizes a YOLOv8n architecture and was trained on a dataset of 12,000 labeled holographic images—comprising 4,000 images each of PD particles (PDPs), *E. coli* (EC), and *P. aeruginosa* (PA). Data augmentation and iterative training with hard data mining techniques enhanced the model's robustness and accuracy. Initially, we used our system to analyze the characteristics of generic PM in sterile PD fluids. The model successfully detected and classified 5,000 individual PDPs, revealing a diverse range of particle morphologies—including spheres, rods, and agglomerates—with sizes predominantly between 1–5 µm. The average concentration of PDPs was estimated to be 61 particles/µL, and the size distribution had a mean diameter of 1.8 µm. This provides valuable baseline information for particulate content in PD fluids, essential for distinguishing contaminants from inherent particles. Furthermore, we assessed the system's capability to differentiate between PDPs and bacteria by spiking PD fluid with high concentrations (>$10^6$ bacteria/mL) of EC or PA. Quantitative evaluation using receiver operating characteristic (ROC) curves demonstrated high areas under the curve (AUCs) nearing 0.98 for both bacterial classifications, reflecting the model's robustness even when particle sizes overlap. At a false positive rate (FPR) of 0.5%, the true positive rate (TPR) reached approximately 80% for EC and 90% for PA. Finally, we validated our DIH system against standard colony-forming unit (CFU) methods using PD fluid spiked with known bacterial concentrations ranging from approximately 100 CFU/mL to 10,000 CFU/mL. Linear regression analysis yielded a high coefficient of determination ($R^2$) of 0.96, indicating a strong linear relationship between DIH counts and CFU counts within the tested concentration range. These results demonstrate that our DIH system can serve as a reliable alternative to traditional culture-based methods, providing rapid and accurate assessments of bacterial contamination in PD fluids.

Our DIH system represents the first real-time analytical platform capable of simultaneously assessing PM and sterility in PD fluids. This innovative approach addresses a critical need in PD treatment and holds significant potential for extension to other liquid injectable drugs in the pharmaceutical industry. Existing imaging-based particle analysis methods, such as Microflow Imaging (MFI) using standard brightfield imaging, experience limitations, including a shallow depth of field and challenges with particles whose refractive index closely matches that of the suspension fluid. MFI is costly and not optimized for routine point-of-care quality assurance. Additionally, MFI has inherent variability in sample loading, which can lead to inconsistent analysis results, and its accuracy can deviate by 25-50% for specific particle types, such as protein aggregates (Narhi et al., 2015; Corvari et al., 2015; Kiyoshi et al., 2019; Fawaz et al., 2023). Holographic systems, such as the xSight by Spheryx, offer an alternative by using holography to measure particles in suspension. However, these systems are also limited, with a sensitivity of $10^3$ particles/mL, and provide only basic classifications by size and refractive index for particles within a 0.5–10 µm range (Boltyanskiy et al., 2022; Spheryx, 2024). Our system offers the advantage of detecting and analyzing particles across a wide size distribution, from submicron levels upwards. This broad detection range is particularly beneficial for pharmaceutical applications where stringent PM standards are essential to ensure product safety and efficacy. Medications that require high purity—such as injectable drugs and vaccines—can greatly benefit from our system's ability to monitor and characterize particles that might compromise quality. In terms of sterility testing, our DIH system surpasses traditional methods by providing real-time, autonomous detection of bacterial contamination. Leveraging deep learning algorithms, the system can be trained to recognize a diverse array of bacterial species and has the potential to detect unknown contaminants (Bravo-Frank et al., 2024). This adaptability not only accelerates the detection process but also enhances the system's robustness in various clinical and industrial settings. The compact and modular design of our system enhances its applicability. It can be deployed as a mobile unit for in-home PD treatments, allowing patients to monitor the sterility of PD fluids in real time and reducing the risk of peritonitis and associated complications. Additionally, the system has the potential to be integrated into pharmaceutical production lines for inline monitoring, providing continuous quality assurance during the manufacturing process. Such integration would improve product safety and increase manufacturing efficiency by enabling immediate detection and remediation of contamination events.

However, the current system operates at a flow rate of 30 μL/min, which may be insufficient for applications requiring higher throughput. To address this limitation, future work could focus on integrating preconcentration techniques. Specifically, our system could analyze the liquid effluent from filter-based preconcentrators, significantly increasing the volume of fluid processed—potentially by several orders of magnitude. This approach would enhance the amount of liquid our system can analyze, thereby improving sensitivity and lowering the detection limit for bacterial contamination. Incorporating preconcentration methods would help us better evaluate the sensitivity of our system, which is currently constrained by the low bacterial counts in the volume of liquid processed. By concentrating the sample, we increase the absolute number of bacteria present in the analyzed volume, enhancing the likelihood of detecting bacteria at lower initial concentrations. This improvement is crucial for detecting bacterial levels below our current threshold, where low bacterial counts relative to the noise floor imposed by false positives reduce detection reliability.

Looking ahead, several avenues can be explored to enhance the system's performance and applicability. Expanding the training dataset to include a broader spectrum of bacteria and particulate types could improve the model's classification accuracy and its ability to detect unknown contaminants. Optimizing the imaging system and data processing pipeline could increase throughput, enabling the analysis of larger sample volumes in shorter times. Additionally, developing advanced deep learning model to differentiate live bacteria from dead ones from holographic signatures (Sanborn et al., 2023; Bravo-Frank et al., 2024) would provide more clinically relevant information for assessing infection risks.

Overall, our DIH system offers a novel, real-time solution for monitoring PM and sterility in PD fluids, with broad applicability in the pharmaceutical industry. By addressing current limitations through system enhancements and integration with preconcentration techniques, we can further improve its performance and utility. This technology holds significant promise for enhancing patient care in PD treatment and improving quality control in pharmaceutical manufacturing.

**Data availability**

The data that support the findings of this study are available on request.

**CRediT authorship contribution statement**

*Nicholas Bravo-Frank:* Writing – review & editing, Writing – original draft, Formal Analysis, Data curation, Conceptualization, Visualization, Investigation
*Nicolas Mesyngier:* Writing – original draft, Formal Analysis, Data Curation, Investigation
*Lei Feng:* Writing – original draft, Investigation
*Jiarong Hong:* Writing – review & editing, Conceptualization, Project Administration

**Declaration of competing interest**

The authors declare that they have no known competing financial interests or personal relationships that could have appeared to influence the work reported in this paper.

**Declaration of generative AI and AI-assisted technologies in the writing process**

During the preparation of this work the authors used ChatGPT 4o to improve the language flow and readability of this work. After using this tool/service, the authors reviewed and edited the content as needed and take full responsibility for the content of the publication.